\documentclass[aps,pra,twocolumn,10pt,letterpaper,groupedaddress]{revtex4-1}
\usepackage{amsmath, amssymb}
\usepackage{xcolor, graphicx}
\usepackage[colorlinks=true,urlcolor=blue,citecolor=blue,linkcolor=magenta]{hyperref}

\begin{document}

\title{Quasi\-states and quasiprobabilities}

\author{J. Sperling}
\affiliation{Clarendon Laboratory, University of Oxford, Parks Road, Oxford OX1 3PU, United Kingdom}

\author{I. A. Walmsley}
\affiliation{Clarendon Laboratory, University of Oxford, Parks Road, Oxford OX1 3PU, United Kingdom}

\date{\today}

\begin{abstract}
	The quasiprobability representation of quantum states addresses two main concerns, the identification of nonclassical features and the decomposition of the density operator.
	While the former aspect is a main focus of current research, the latter decomposition property has been studied less frequently.
	In this contribution, we introduce a method for the generalized expansion of quantum states in terms of so-called quasi\-states.
	In contrast to the quasiprobability decomposition through nonclassical distributions and pure-state operators, our technique results in classical probabilities and nonpositive semidefinite operators, defining the notion of quasi\-states, that carry the information about the nonclassical characteristics of the system.
	Therefore, our method presents a complementary approach to prominent quasiprobability representations.
	We explore the usefulness of our technique with several examples, including applications for quantum state reconstruction and the representation of nonclassical light.
	In addition, using our framework, we also demonstrate that inseparable quantum correlations can be described in terms of classical joint probabilities and tensor-product quasi\-states for an unambiguous identification of quantum entanglement.
\end{abstract}

\maketitle

\section{Introduction}

	Quantum systems can exhibit properties which have no analog in the classical realm, opening possibilities for technological innovations beyond the limitations posed by classical physics \cite{NC00}.
	For assessing the quantumness of a state, the bisection of a system into a classically accessible domain and a genuinely quantum-mechanical sector has become one of the major challenges of current research.
	A successful approach to performing such a desired separation are quasiprobability distributions, such as the Wigner-Weyl function \cite{W32,W27}.
	Within this framework, a state is identified to be nonclassical when the corresponding quasiprobability does not have the properties of a classical probability distribution.

	Beyond this nonclassicality aspect, another purpose of quasiprobabilities is the full characterization of a quantum state, for example, using the Glauber-Sudarshan representation to describe quantized harmonic oscillators \cite{G63,S63}.
	As such, the description of a density operator in terms of quasiprobabilities and classical reference states, e.g., coherent states, is an equally important feature of quasiprobabilities.
	However, despite having such a vital impact on the state's representation, the density operator expansion is, by far, less frequently considered.
	This is at least partially a result of a missing framework which focuses on the decomposition of density operators.
	Thus, the main objective of this contributions is to devise such a missing methodology that naturally leads to the useful and complementary concept of quasi\-states.

	The notion of quasiprobabilities has become one of the main hallmarks for certifying quantum features since it allows us to study resources for practical tasks.
	For instance, negative quasiprobabilities have been identified as indicators for the state's usefulness in performing quantum computation protocols \cite{F11,VFGE12}.
	Moreover, negativities in quasiprobabilities can be related to a broad range of concepts of nonclassicality, such as contextuality \cite{S08} and quantum entanglement \cite{DMWS06}.
	In addition, quasiprobabilities can be helpful to determine the actual probabilities of measurement outcomes \cite{PWB15,KSVS18}.

	Furthermore, quasiprobabilities can be defined for many relevant physical systems, such as continuous-variable harmonic oscillators \cite{G63,S63} or discrete-variable angular-momentum-type degrees of freedom \cite{AD71,A81}.
	Typically, the underlying quasiprobabilities are obtained by generalizing a classical phase space to the quantum domain; see Ref. \cite{BM98} for an early study and Ref. \cite{TESMN16} for a recent generalization.
	For example, quasiprobabilities in finite-dimensional systems can be introduced in this manner, cf., e.g., Refs. \cite{LP98,KMR06,PB11,SW18}.
	With regard to the quantum state representation, this typically requires us to identify pure reference states which are compatible with the properties of classical physics \cite{P72}.
	Based on such classical reference states, a general quasiprobability decomposition of a quantum state can be constructed \cite{SW18}.
	Examples relevant for quantum technologies include the formulation of entanglement quasiprobabilities \cite{STV98,SV09}, which become negative for quantum correlated states.

	Once a quasiprobability representation is constructed, it can be further generalized.
	Again, pioneering applications can be found in quantum optics, where the Glauber-Sudarshan and Wigner-Weyl quasiprobabilities can be unified in terms of so-called $s$-parametrized quasiprobabilities \cite{CG69,CG69prime}, which also include the well-known Husimi-Kano distributions \cite{H40,K65}.
	This unification is achieved by a convolution of the quasiprobability with a Gaussian function and can be further generalized \cite{AW68,AW70,AW70prime}, including non-Gaussian scenarios \cite{K66,KV10}.
	A successful application of the $s$-parameter approach in a finite-dimensional system has been reported as well \cite{RMG05}.
	In addition, joint non-Gaussian quasiprobabilities for multiple optical modes and points in time have been introduced to study quantum correlated light \cite{ASV13,KVS17}.

	Although such formal aspects led to profound insights into quantum physics when compared to classical statistical theories (see, e.g., \cite{M49,K13}), the method of quasiprobabilities is also of experimental importance.
	The most frequent implementation is the identification of nonclassical light in terms of negative Wigner-Weyl functions; see Refs. \cite{SBRF93,HSRHMSS16} for early and recent examples.
	While the Glauber-Sudarshan distribution can be highly singular or even ambiguous \cite{KMC65,BV87,S16}, the reconstruction of this function is feasible in certain experiments as well \cite{KVPZB08}.
	Moreover, the non-Gaussian generalization of this distribution is always experimentally accessible, such as reported for squeezed light \cite{KVHS11}.
	Even imperfect detection schemes can be employed for the reconstruction of nonclassical phase-space distribution of light \cite{BTBSSV18}.
	Conversely, a phase-space representation of the detector can experimentally verify quantum properties of the detection device used \cite{LFCPSRWPW09}.
	Another approach \cite{RMH10,MIMSRH13} enables an experimental reconstruction of phase-space distributions and entirely circumvents a detector characterization by analyzing certain data patterns \cite{CKS14}.
	Beyond quantum light, equally insightful is the characterization of matter systems using quasiprobabilities, such as demonstrated for motional states of ions \cite{LMKMIW96} and large atomic ensembles \cite{MZHCV15}.
	It is also worth mentioning that quantum correlations in composite hybrid systems are also accessible via generalized quasiprobabilities \cite{SAWV17,ASCBZV17}.

	Therefore, quasiprobabilities present a highly successful and versatile approach to identifying nonclassical properties of quantum states in theory and experiment.
	Still, as outlined above, another important aspect is the quantum state decomposition.
	Yet, in contrast to the vast number of examples for the nonclassicality certification, studies of the decomposition properties are rarely done;
	a gap we aim to close in this article.

	Based on the decomposition of states in terms of quasiprobabilities, we introduce a generalized expansion of quantum states in terms of quasi\-states.
	This method is complementary to the previously known approaches in which a density operator is decomposed in terms of classical reference states and a negative quasiprobability density.
	In our general method, we can find a decomposition which results in a nonnegative (i.e., classical) probability density by overcoming the usage of physical reference states in the decomposition.
	This naturally establishes the concept of quasi\-states, which then become the relevant objects for certifying the different kinds of nonclassicality.
	The usefulness of this change of perspective from quasiprobabilities to quasi\-states is studied in detail.
	One practical application relates to the experimental reconstruction of density operators.
	Moreover, as quasiprobabilities are most frequently applied in optical systems, we also perform a detailed analysis of quasi\-states that correspond to prominent phase-space representations of light.
	Finally, we investigate how quantum entanglement can be uniquely characterized via classical joint probability distributions when employing quasi\-states.
	Therefore, our approach offers a toolbox for the characterization and decomposition of quantum states.

	This article is structured as follows.
	In Sec. \ref{sec:Framework}, we discuss the general methodological framework.
	An application to the quantum state reconstruction is developed in Sec. \ref{sec:StateRecon}.
	Phase-space based quasi\-states for quantized light are studied in Sec. \ref{sec:NclLight}.
	In Sec. \ref{sec:Correlations}, quantum correlations are analyzed within the proposed framework.
	Finally, concluding discussions are presented in Sec. \ref{sec:SumConcl}.

\section{Conceptual framework}\label{sec:Framework}

	To characterize the state of a quantum system, a representation of the density operator is required.
	In general, a state decomposition consists of a family of pure states, defining a set $\mathcal S$, and corresponding expansion coefficients.
	Then the density operator takes the form
	\begin{align}
		\label{eq:InitialDecomposition}
		\hat\rho=\int_{\mathcal S} d\psi\, P(\psi) |\psi\rangle\langle\psi|,
	\end{align}
	where $d\psi$ indicates an integral over the volume of $\mathcal S$ and $P$ is a correspondingly defined density.
	In the case of a discrete decomposition, the integral may be replaced by a sum.
	Note that we deliberately choose not to specify the set $\mathcal S$ to keep our treatment as broadly applicable as possible.

	Arguably the most well-known example of a representation \eqref{eq:InitialDecomposition} is the spectral decomposition.
	In this case, $\mathcal S$ is the set of eigenstates and the probability mass function $P$ returns the corresponding eigenvalues.
	Other examples for a decomposition \eqref{eq:InitialDecomposition} are studied in the continuation of this work and have applications, for example, in quantum optics and quantum information.
	It is worth mentioning that a general method to construct quasiprobabilities $P$ for a given set $\mathcal S$ has been recently devised \cite{SW18}.
	In that approach, it is ensured that $P$ is a classical probability distribution when the state is in the convex hull $\mathrm{conv}\{|\psi\rangle\langle\psi|:\psi\in\mathcal S\}$, which is a nontrivial task as the decomposition \eqref{eq:InitialDecomposition} is typically not unique \cite{SW18}.

	For our purpose, we may generalize the above treatment.
	Specifically, a convolution kernel $K:\mathcal S\to\mathcal S$ is introduced together with the kernel $K^{-1}$ for the corresponding deconvolution.
	Then we get an equivalent decomposition as
	\begin{align}
		\label{eq:ModifiedDecomposition}
		\hat\rho=\int_{\mathcal S} d\chi\, P_K(\chi)\hat\Delta_{K}(\chi).
	\end{align}
	Therein we have a modified density $P_K$ and a modified family of operators $\hat\Delta_{K}$,
	\begin{align}
		\label{eq:QuasiProbability}
		P_K(\chi)=&\int_{\mathcal S} d\psi\, P(\psi)K(\psi,\chi),
		\\
		\label{eq:QuasiState}
		\hat\Delta_{K}(\chi)=&\int_{\mathcal S} d\psi'\, K^{-1}(\chi,\psi') |\psi'\rangle\langle\psi'|.
	\end{align}
	Using the properties that the operations used are inverse to one another, $\delta(\psi,\psi')=\int_{\mathcal S} d\chi K(\psi,\chi)K^{-1}(\chi,\psi')$ with $\delta$ being the Dirac distribution, one can directly see that Eqs. \eqref{eq:InitialDecomposition} and \eqref{eq:ModifiedDecomposition} describe the same density operator $\hat \rho$.

	In general, the distribution $P_K$ in Eq. \eqref{eq:QuasiProbability} is not a probability density, even if $P$ was.
	For such generalized distributions, the name quasiprobabilities was established.
	As discussed in the Introduction, such quasiprobability densities cover a wide range of applications, mainly for purpose of identifying quantum features.
	In close analogy to the notion of a quasiprobability, we are going to demonstrate that the operators $\hat\Delta_K$ in Eq. \eqref{eq:QuasiState} are, in general, not physical density operators.
	Consequently, we refer to such operators as quasi\-states.

	A main focus of previous research was devoted to characterizing quasiprobabilities.
	Often, the idea of a decomposition of the quantum state [cf. Eqs. \eqref{eq:InitialDecomposition} and \eqref{eq:ModifiedDecomposition}] in terms of such distributions was neglected.
	In particular, a general characterization of the distinctive features of quasi\-states and their applications beyond being a mathematical tool does not exist.
	For this reason, we are going to perform the missing comprehensive investigation of quasi\-states as defined in Eq. \eqref{eq:QuasiState} and discuss useful applications of such operators.

\section{State reconstruction}\label{sec:StateRecon}

	One interesting application of quasi\-states are state reconstruction protocols as we show in this section.
	The following considerations are based on a recent work \cite{KSVS18} in which the dual representation of measurement operators has been introduced.
	Here we demonstrate how this method relates to quasi\-states and can be used for the general reconstruction of density operators.

\subsection{Dual representation}

	Let us briefly revisit the findings in Ref. \cite{KSVS18} with regards to our method.
	Suppose $\{\hat\Pi(j)\}_{j\in\mathcal S}$ is a positive operator-valued measure (POVM).
	As a complementary set of operators, the notion of contravariant operator-valued measured (COVM) was introduced.
	This defines a set $\{\hat\Gamma(j')\}_{j'\in\mathcal S}$ that satisfies the orthonormality relation
	\begin{align}
		\label{eq:Duality}
		\mathrm{tr}\big[\hat\Gamma(j')\hat\Pi(j)\big]=\delta(j',j).
	\end{align}
	This means that the COVM represent dual basis operators to the measured POVM.

	In contrast to the POVM, COVM operators are not necessarily positive semidefinite.
	However, they are of particular interest, for example, when considering imperfections in the measurement process \cite{KSVS18}.
	The construction of COVM operators is based on a kernel defined via the elements
	\begin{align}
		\label{eq:Metric}
		K(l,j)=\mathrm{tr}\big[\hat\Pi(l)\hat\Pi(j)\big].
	\end{align}
	Specifically, it was shown that
	\begin{align}
		\label{eq:COVM}
		\hat\Gamma(j')=\sum_{l\in\mathcal S}K^{-1}(j',l)\hat\Pi(l),
	\end{align}
	for the discrete case.
	In the following, let us explore the relation between the COVM and quasi\-states.

	For simplicity, we assume that the considered complex Hilbert space is finite dimensional, $\dim_{\mathbb C}\mathcal H=d<\infty$.
	The set of Hermitian operators over $\mathcal H$ is a real valued space with the dimension $\dim_{\mathbb R}\mathrm{Herm}(\mathcal H)=d^2$.
	Further, rather than restricting to a POVM for a single observable as done previously, we consider an informationally complete set $\{\hat\Pi(\psi)\}_{\psi\in\mathcal S}$, with
	\begin{align}
		\label{eq:POVM}
		\hat\Pi(\psi)=|\psi\rangle\langle\psi|,
	\end{align}
	which guarantees that a reconstruction of the full quantum state is possible and implies the cardinality $|\mathcal S|=d^2$.
	However, this also means that not all elements of $\mathcal S$ can be pairwise orthogonal vectors.
	See Refs. \cite{P77,APS04,RBSC04} for introductions to informationally complete measurements.

	Because of the properties of the now studied POVM, we can expand the quantum state as given in Eq. \eqref{eq:InitialDecomposition},
	\begin{align}
		\hat\rho=\sum_{\psi\in\mathcal S}P(\psi) \hat\Pi(\psi).
	\end{align}
	It is important to mention that $P(\psi)$, in general, does not correspond to the probability of the measurement of $\hat\Pi(\psi)$, i.e., $\mathrm{tr}[\hat\rho\hat\Pi(\psi)]\neq P(\psi)$.
	However, we can employ the same formalism as discussed above, cf. Eqs. \eqref{eq:Metric} and \eqref{eq:COVM}, to expand the state as \cite{KSVS18}
	\begin{align}
		\label{eq:Reconstruction}
		\hat\rho=\sum_{\chi\in\mathcal S}\rho(\chi) \hat\Gamma(\chi).
	\end{align}
	Now we can identify the coefficient with the measurement probability, $\rho(\psi')=\mathrm{tr}[\hat\rho\hat\Pi(\psi')]$ for all $\psi'\in\mathcal S$, because of relation \eqref{eq:Duality}.
	In other words, the expansion in Eq. \eqref{eq:Reconstruction} can be used to directly reconstruct the density operator in terms of measurement outcomes and dual operators.

	Let us specifically relate the found reconstruction with the generalized decomposition \eqref{eq:ModifiedDecomposition}.
	Using the definitions \eqref{eq:Metric} and \eqref{eq:POVM}, we find that the kernel under study is given by the $d^2\times d^2$ Gram-Schmidt matrix
	\begin{align}
		\label{eq:GramSchmidt}
		K=\left[|\langle\psi|\chi\rangle|^2\right]_{\psi,\chi\in\mathcal S}.
	\end{align}
	Recall that information completeness implies that this matrix is a bijection, thus invertible.
	Further, as a result of Eqs. \eqref{eq:QuasiState} and \eqref{eq:COVM}, the thought-after quasi\-states are identical to the COVM elements,
	\begin{align}
		\hat\Delta_K(\chi)=\hat\Gamma(\chi).
	\end{align}
	Consequently, it also holds true that $P_K(\chi)=\rho(\chi)$; see Eqs. \eqref{eq:ModifiedDecomposition} and \eqref{eq:Reconstruction}.
	Therefore, quasi\-states allow for the direct reconstruction of the density operator under study.

	It is worth mentioning that imperfections in any measurement, meaning that $\hat\Pi(\psi)$ is not a pure-state projector, can be treated similarly to the thorough discussion for a single observable carried out in Ref. \cite{KSVS18}.
	Furthermore, if the POVM does not allow for a full reconstruction or is over-complete (e.g., because $|\mathcal S|\neq d^2$), then the state in the spanned subspace can be reconstructed or a pseudo-inversion of $K$ can be performed \cite{KSVS18}.

\subsection{Example}

\begin{figure}[tb]
	\includegraphics[width=\columnwidth]{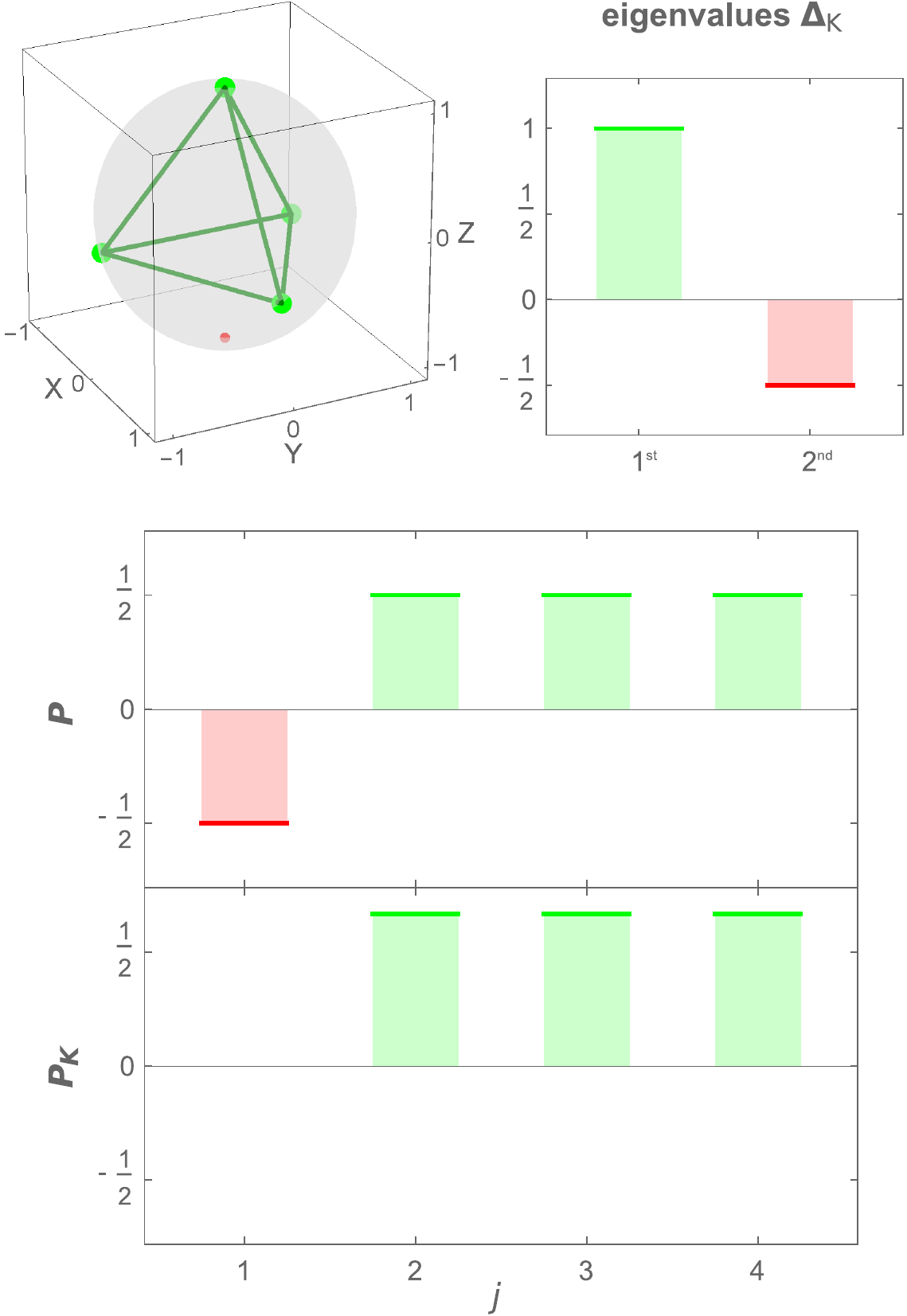}
	\caption{(Color online)
		The elements of $\mathcal S$ given in Eq. \eqref{eq:QubitProjectors} in a tetrahedron configuration (larger green bullets) are shown in the top-left Bloch-sphere representation.
		The smaller red bullet indicates the state $\hat\rho=|0\rangle\langle 0|$ to be reconstructed.
		The top-right panel shows the two eigenvalues of the resulting quasi\-states $\hat\Delta_K(\psi_j)$, being identical for $j=1,\ldots,4$.
		For the reconstruction of the state under study, the bottom panels depict the expansion coefficients $P(\psi_j)$ and $P_K(\psi_j)$ to be used with $|\psi_j\rangle\langle\psi_j|$ and $\hat\Delta_K(\psi_j)$ [Eqs. \eqref{eq:InitialDecomposition} and \eqref{eq:ModifiedDecomposition}], respectively.
	}\label{fig:Reconstruction}
\end{figure}

	To further discuss the above findings, we study the specific example of a qubit system ($d=2$), defined via the orthonormal Hilbert space basis $\{|0\rangle,|1\rangle\}$.
	The four rank-one projection operators $|\psi_j\rangle\langle\psi_j|$ in Eq. \eqref{eq:POVM}, also defining the set $\mathcal S$, can be chosen according to the states
	\begin{align}
		\nonumber
		|\psi_1\rangle=|1\rangle,
		|\psi_2\rangle=&\frac{\sqrt2 |0\rangle+|1\rangle}{\sqrt 3},
		|\psi_3\rangle=\frac{\sqrt2 w|0\rangle+|1\rangle}{\sqrt 3},
		\\\label{eq:QubitProjectors}
		\text{and }
		|\psi_4\rangle=&\frac{\sqrt2 w^2|0\rangle+|1\rangle}{\sqrt 3},
	\end{align}
	where $w=\exp[2\pi i/3]$; see the top-left panel in Fig. \ref{fig:Reconstruction} for their geometric Bloch-sphere representation.
	This allows us to construct $K$ via Eq. \eqref{eq:GramSchmidt} and compute its inverse,
	\begin{align}
		K=\frac{1}{3}\begin{bmatrix}
			3 & 1 & 1 & 1
			\\
			1 & 3 & 1 & 1
			\\
			1 & 1 & 3 & 1
			\\
			1 & 1 & 1 & 3
		\end{bmatrix}
		\!,\,
		K^{-1}=\frac{1}{4}\begin{bmatrix}
			5 & -1 & -1 & -1
			\\
			-1 & 5 & -1 & -1
			\\
			-1 & -1 & 5 & -1
			\\
			-1 & -1 & -1 & 5
		\end{bmatrix}
		\!.
	\end{align}
	This then yields the quasi\-states as
	\begin{align}
		\label{eq:QubitQuasistates}
		\hat \Delta_K(\psi_1)=&|1\rangle\langle1|-\frac{1}{2}|0\rangle\langle 0|
		\text{ and}
		\\\nonumber
		\hat \Delta_K(\psi_j)=&
		\frac{
			|0\rangle\langle0|
			+\sqrt2w^{j-2}|0\rangle\langle1|
			+\sqrt2w^{\ast (j-2)}|1\rangle\langle0|
		}{2},
	\end{align}
	for $j=2,3,4$.
	As we can see from the top-right panel in Fig. \ref{fig:Reconstruction}, these quasi\-states do not resemble physical quantum states because of the negative eigenvalue.

	Let us now consider the specific state $\hat\rho=|0\rangle\langle 0|$ for the reconstruction.
	Then, the measured and, therefore, nonnegative expansion coefficients are
	\begin{align}
		P_K(\psi_j)=\mathrm{tr}(\hat\rho|\psi_j\rangle\langle\psi_j|)=\langle\psi_j|\hat\rho|\psi_j\rangle,
	\end{align}
	using the projectors defined via the states in Eq. \eqref{eq:QubitProjectors}.
	In addition, the corresponding COVM elements, being the quasi\-states in Eq. \eqref{eq:QubitQuasistates}, result in the expansion coefficients $P(\psi_j)=\mathrm{tr}[\hat\rho\hat\Delta_K(\psi_j)]$, which resembles a quasiprobability distribution.
	The bottom part in Fig. \ref{fig:Reconstruction} shows the values of $P_K$ and $P$ for the state $\hat\rho=|0\rangle\langle 0|$.

	Using the nonnegative $P_K$ (see Fig. \ref{fig:Reconstruction}), we can now directly confirm that the density operator $\hat\rho=|0\rangle\langle 0|$ is expanded as described in Eq. \eqref{eq:ModifiedDecomposition} while using the quasi\-states in Eq. \eqref{eq:QubitQuasistates}, $\hat\rho=\sum_{j=1,\ldots,4}P_K(\psi_j)\hat\Delta_K(\psi_j)$.
	The possibly negative elements of $P$ also allow for the decomposition of the state, $\hat\rho=\sum_{j=1,\ldots,4}P(\psi_j)|\psi_j\rangle\langle\psi_j|$.
	However, $P$ does not correspond to a directly measured quantity.
	Thus, quasi\-states provide a beneficial tool to reconstruct density operators using measured quantities.

	Let us also briefly comment on the extension of our approach to continuous-variable systems, $d=\infty$.
	In fact, already in Refs. \cite{R96,LMKRR96}, a method was introduced to expand the state of a single-mode radiation field in terms of so-called pattern functions.
	Using the here developed framework, we can directly identify these pattern functions with the wave functions of the corresponding quasistates.
	Beyond this reconstruction approach, we study the expansion of nonclassical states of light in the following section.

\section{Quantum Light}\label{sec:NclLight}

	We now demonstrate the application of quasi\-states for the description of nonclassical quantum states.
	In particular, we consider a harmonic oscillator, which can, for example, describe a quantized radiation mode.
	To describe the system's nonclassicality, the prominent Glauber-Sudarshan representation has been developed \cite{G63,S63},
	\begin{align}
		\hat\rho=\int_{\mathbb C}d\alpha\,P(\alpha)|\alpha\rangle\langle\alpha|.
	\end{align}
	Here this decomposition implies that we identify the set $\mathcal S$ with the complex coherent amplitudes $\alpha\in\mathbb C$, defining the coherent states $|\alpha\rangle$.
	For a thorough introduction to quantum optics in phase space, see Ref. \cite{VW06}.
	A state is referred to as nonclassical if $P$ does not exhibit the properties of a classical probability distribution.
	In the following, we begin our considerations of quasi\-states for nonclassical light in the Fourier picture and then proceed to prominent examples of quasiprobabilities and their dual representation via quasi\-states.

\subsection{Fourier representation}

	The characteristic function $\Phi(\beta)$ is the Fourier transform $F$ of a distribution, which is accessible via the kernel
	\begin{align}
		F(\alpha,\beta)=\exp\left(
			\beta\alpha^\ast-\beta^\ast\alpha
		\right),
	\end{align}
	i.e., in our notation, $P_{F}=\Phi$.
	The inverse Fourier transform, defined via $F^{-1}(\beta,\alpha)=F(\alpha,\beta)^\ast/\pi^2$, results in the corresponding quasi\-states [Eq. \eqref{eq:QuasiState}],
	\begin{align}
		\hat\Delta_{F}(\beta)=\int_{\mathbb C}d\beta\,\frac{\exp\left(\beta^\ast\alpha-\beta\alpha^\ast\right)}{\pi^2}|\alpha\rangle\langle\alpha|.
	\end{align}
	For the evaluation of the integral, it is useful to represent the coherent-state projectors in terms of normally ordered exponential of creation ($\hat a^\dag$) and annihilation ($\hat a$) operators, which reads $|\alpha\rangle\langle\alpha|={:}\exp(-[\hat a-\alpha]^\ast[\hat a-\alpha]){:}$ \cite{VW06}.
	This representation yields
	\begin{align}
		\hat\Delta_{F}(\beta)=\frac{\exp(-|\beta|^2)}{\pi}{:}\hat D(-\beta){:},
	\end{align}
	where $\hat D(\gamma)=\exp(\gamma\hat a^\dag-\gamma^\ast\hat a)=\exp(-|\gamma|^2/2){:}\hat D(\gamma){:}$ is the unitary displacement operator.
	Therefore, the quasi\-states $\hat\Delta_{F}(\beta)$ correspond (up to a rescaling) to unitary operators in the case that the convolution kernel is the Fourier transformation.
	Note that instead of the Fourier transform, the Laplace transform can be advantageous in certain scenarios as well \cite{SVA16}.

	One application, which also connects to the previous section, is the state reconstruction via balanced homodyne detection.
	For this purpose, let us recall that the characteristic function of the Wigner-Weyl distribution takes the form \cite{VW06}
	\begin{align}
		\Phi(\beta)e^{-|\beta|^2/2}=\langle\hat D(\beta)\rangle
		=\langle\exp\left(
			i|\beta|\hat x[\pi/2-\arg\beta]
		\right)\rangle,
	\end{align}
	where we used the quadrature operator $\hat x[\varphi]=e^{i\varphi}\hat a+e^{-i\varphi}\hat a^\dag$.
	This means for a balanced homodyne measurement, providing a data set $\{x_j[\varphi_j]\}_{j=1,\ldots,N}$ of quadratures $x_j$ for the phase values $\varphi_j$, we can approximate
	\begin{align}
		\label{eq:BHD}
	\begin{aligned}
		\hat\rho
		=&\int_{\mathbb C}d\beta\,\Phi(\beta)\hat\Delta_{F}(\beta)
		\\=&
		\int_{-\infty}^\infty dr\int_{-\pi/2}^{\pi/2} d\varphi\,\frac{\pi}{\pi}
		e^{r^2/2}\langle e^{ir\hat x[\varphi]}\rangle\hat\Delta_{F}(ire^{-i\varphi})
		\\\approx&\int_{-\infty}^\infty dr\, e^{r^2/2} \frac{\pi}{N}\sum_{j=1}^N e^{irx_j[\varphi_j]}\hat\Delta_{F}(ire^{-i\varphi_j}),
	\end{aligned}
	\end{align}
	which corresponds to a sampling approach as a consequence of a quasi\-state representation [see specifically the first line of Eq. \eqref{eq:BHD} and the use of $\hat\Delta_{F}$].
	In fact, further analyses show that this approach is indeed equivalent to established reconstruction methods bases on balanced homodyne detection and using pattern functions \cite{R96,LMKRR96,LR09}.
	In addition, it is worth emphasizing that the phases $\varphi$ have to be uniformly distributed in the interval $[-\pi/2,\pi/2]$ to ensure an optimal sampling \cite{ASVKMH15}.

\begin{figure*}
	\includegraphics[width=\textwidth]{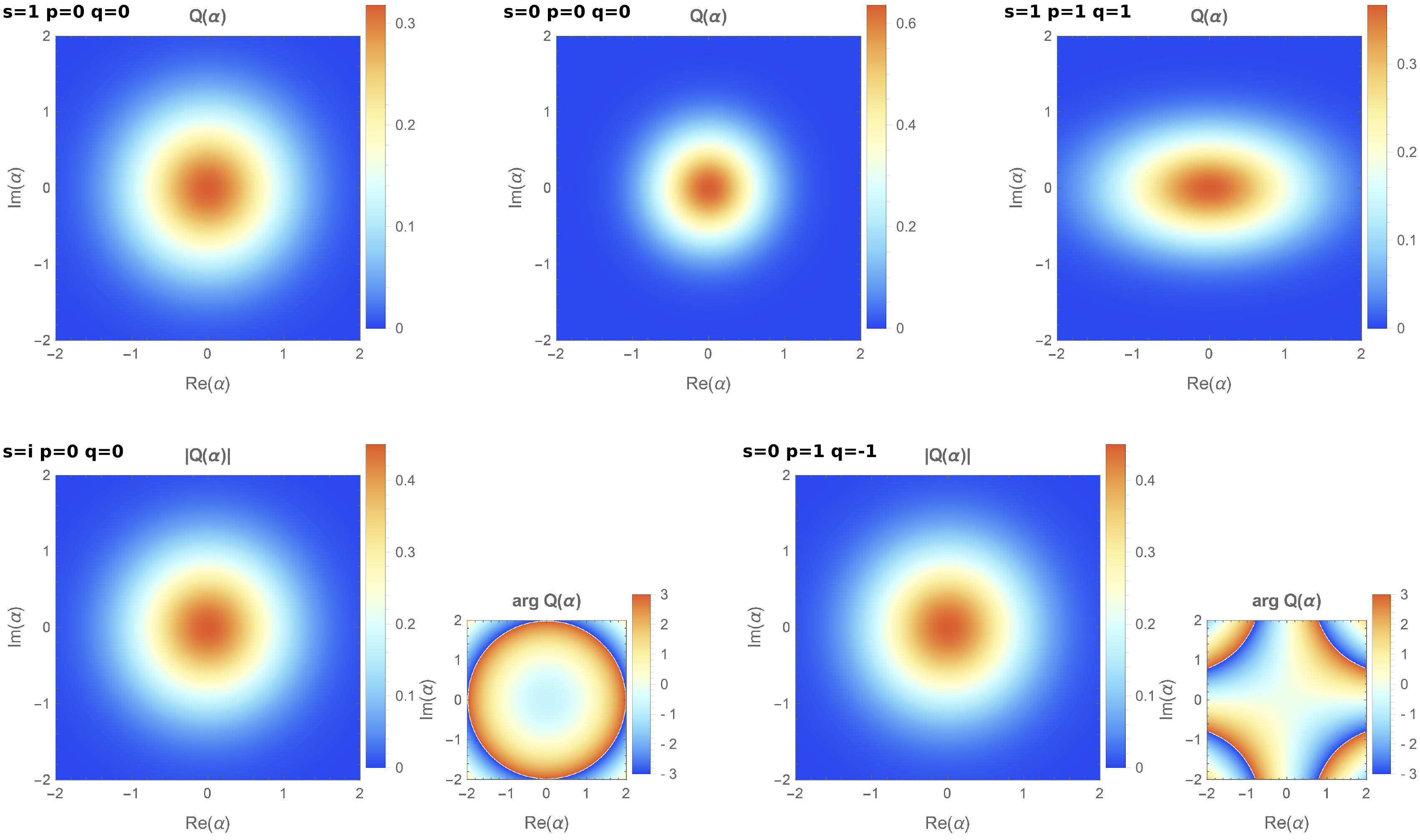}
	\caption{(Color online)
		The $Q$ function [Eq. \eqref{eq:Qfct}] of the quasi\-state [Eq. \eqref{eq:QOQS}] for different parameters $s$, $p$, and $q$.
		The top-left plot corresponds to a coherent (vacuum) state.
		The top-center plot shows the maximally singular quasi\-state \cite{S16}.
		The top-right plot includes, in addition, squeezing.
		The bottom row depicts cases in which the $Q$ function becomes complex;
		the larger and corresponding smaller plots show the amplitude $|Q|$ and phase $\arg Q$, respectively.
		Note the difference in the phases when comparing the left and right plots.
	}\label{fig:QO}
\end{figure*}

\subsection{Generalized optical quasi\-states}

	Translational types of convolutions in the original space, e.g., $P_{\kappa}(\alpha')=\int_{\mathbb C}d\alpha\,P(\alpha)\kappa(\alpha'-\alpha)$, take a product form in the Fourier domain.
	In this scenario, we can consider the representation
	\begin{align}
		\hat\rho=\int_{\mathbb C}d\beta\,\left[\Phi(\beta)\Omega(\beta)\right]\left[\frac{e^{-|\beta|^2}}{\pi\Omega(\beta)}{:}\hat D(-\beta){:}\right],
	\end{align}
	for a kernel $\Omega$.
	The inverse Fourier transform gives the distribution $P_{F^{-1}\Omega F}$ as well as the quasi\-states
	\begin{align}
		\hat\Delta_{F^{-1}\Omega F}(\alpha)=\int_{\mathbb C}d\beta\,e^{\beta\alpha^\ast-\beta^\ast\alpha}\frac{e^{-|\beta|^2}}{\pi\Omega(\beta)}{:}\hat D(-\beta){:}.
	\end{align}
	Such operators have been extensively studied in connection to operator orderings and their resulting quasiprobabilities \cite{CG69,CG69prime,AW68,AW70,AW70prime}.
	Here, however, let us analyze them based on their own merit as quasi\-states.

	Of particular interest are the Gaussian functions
	\begin{align}
		\label{eq:GaussFilter}
		\Omega(\beta)=\exp\left(
			-\frac{1-s}{2}|\beta|^2-\frac{p}{4}\beta^2-\frac{q}{4}\beta^\ast
		\right),
	\end{align}
	where $|p|=|q|$ \cite{AW68,AW70,AW70prime}.
	In Table \ref{tab:sParameters}, specific examples and their connection to known quasiprobabilities are given.
	Evaluating the integrals as demonstrated in Appendix \ref{app:ExpOp}, we get the quasi\-states in normal ordering as
	\begin{align}
		\label{eq:QOQS}
		\hat\Delta_{F^{-1}\Omega F}(0)
		=\frac{2\,{:}\exp\left(
			\frac{-2[1+s]\hat a^\dag\hat a +q\hat a^{\dag 2}+p\hat a^2}{(1+s)^2-pq}
		\right){:}}{\sqrt{(1+s)^2-pq}}.
	\end{align}
	Note that because of the convolution property, we have $\hat\Delta_{F^{-1}\Omega F}(\alpha)=\hat D(\alpha)\hat\Delta_{F^{-1}\Omega F}(0)\hat D(\alpha)^\dag$ for the displaced quasi\-states.
	In Fig. \ref{fig:QO}, several examples of the phase-space $Q$-function representation,
	\begin{align}
		\label{eq:Qfct}
		Q(\alpha)=\frac{\langle\alpha|\hat\Delta_{F^{-1}\Omega F}(0)|\alpha\rangle}{\pi},
	\end{align}
	are shown for different parameters that define the quasi\-states in Eq. \eqref{eq:QOQS}.
	It is also worth mentioning that the $Q$ function is the Husimi-Kano distribution.

\begin{table}[b]
	\caption{
		Parameters [cf. Eq. \eqref{eq:GaussFilter}] for prominent quasiprobability distributions frequently used in quantum optics.
	}\label{tab:sParameters}
	\begin{tabular}{p{5cm} p{.2cm} c p{.2cm} c p{.2cm} c  p{.2cm}}
		\hline\hline
		quasiprobability && $s$ && $p$ && $q$ &
		\\\hline
		Glauber-Sudarshan \cite{G63,S63} && $1$ && $0$ && $0$ &
		\\
		Wigner-Weyl \cite{W27,W32} && $0$ && $0$ && $0$ &
		\\
		Husimi-Kano \cite{H40,K65} && $-1$ && $0$ && $0$ &
		\\
		Agarwal-Wolf  \cite{AW68,AW70,AW70prime} && $0$ && $\pm1$ && $\mp 1$ &
		\\\hline\hline
	\end{tabular}
\end{table}

	Interestingly, the considered class of quasi\-states, given in Eq. \eqref{eq:QOQS}, can be related to squeezed versions of thermal-like quasi\-states,
	\begin{align}
		\label{eq:QOQSdecomposition}
	\begin{aligned}
		\hat\Delta_{F^{-1}\Omega F}(0)
		=&
		\exp([\zeta^\ast\hat a^2-\zeta\hat a^{\dag2}]/2)
		\\
		&\times
		\frac{1}{\bar n+1}\sum_{n\in\mathbb N}\left(\frac{\bar n}{\bar n+1}\right)^n|n\rangle\langle n|
		\\
		&\times
		\exp([\zeta^\ast\hat a^2-\zeta\hat a^{\dag2}]/2)^\dag,
	\end{aligned}
	\end{align}
	where $\bar n\in\mathbb C$ generalizes the notion of a mean thermal photon number and $\zeta$ is the squeezing parameter.
	A full and exact analysis is provided in Appendix \ref{app:ExpOp}, and it provides the relations between the different parameter sets as
	\begin{align}
		\label{eq:ParameterRelations}
		\zeta=e^{i\arg\tau}\mathrm{artanh}\,|\tau|
		\quad\text{and}\quad
		\bar n=\frac{\omega}{1-\omega},
	\end{align}
	with
	\begin{align}
		\label{eq:ParameterRelationsPlus}
		\omega=\frac{r-1}{r+1},\,
		|\tau|^2=\frac{s-r}{s+r},\,
		\text{and}\,
		\frac{\tau}{\tau^\ast}=\frac{q}{p},
	\end{align}
	where $r=\sqrt{s^2-pq}$.
	Recall that we require $|q|=|p|$.
	Also note that the squeezing operation is unitary which implies that we have the eigenvalues $(1-\omega)\omega^n$, following a geometric distribution.
	In the following, let us study the resulting quasi\-states in more detail.

\subsubsection{$s$-Parameterized quasi\-states}

	Arguably the most frequently used choice of parameters is obtained for $p=q=0$, likewise $\zeta=\tau=0$.
	This choice results in the $s$-parametrized quasiprobability distributions $P_{F^{-1}\Omega F}$.
	Here, we additionally find the quasi\-states $\hat\Delta_{F^{-1}\Omega F}$.
	In particular, those quasi\-states have the eigenvalues
	\begin{align}
		\label{eq:sParaEvals}
		(1-\omega)\omega^n
		=\frac{\bar n^{n}}{(\bar n+1)^{n+1}}
		=\frac{2}{1+s}\left(
			-\frac{1-s}{1+s}
		\right)^n
	\end{align}
	for $n\in\mathbb N$; see Eq. \eqref{eq:QOQSdecomposition}.

	For the parameters $-1<s<1$, the eigenvalues \eqref{eq:sParaEvals} of the quasi\-states are negative for odd $n$, certifying that these operators are only accessible in terms of our generalization of the notion of a physical state.
	In the limit $s\to1$, we get the eigenvalue $1$ for $n=0$ and $0$ for $n>0$, defining the vacuum state.
	Its $Q$ function is shown on the left in the top row of Fig. \ref{fig:QO}, and the corresponding quasiprobability distribution is the original Glauber-Sudarshan distribution.
	Furthermore, the case $s=0$ interestingly describes an operator with the maximally possible singularities a Glauber-Sudarshan distribution can exhibit \cite{S16};
	see also the center plot in the top row of Fig. \ref{fig:QO} for its $Q$ function.
	There, we observe that the spread (i.e., variance) in phase space is reduced in all directions when compared to the vacuum state, further highlighting that it is a genuine quasi\-state that is incompatible with the uncertainty relation which holds true for any physical state.
	The underlying phase-space distribution $P_{F^{-1}\Omega F}$ is the Wigner-Weyl distribution.
	Finally, the quasi\-states for the Husimi-Kano distribution are obtained in the limit $s\to-1$, and its corresponding $Q$ function is described by a singular Dirac distribution.

	Beyond real-valued $s$ parameters, we can additionally consider complex values as well \cite{VW06}.
	Then, the eigenvalues \eqref{eq:sParaEvals} are complex too.
	The example $s=i$ is shown on the left of the bottom row in Fig. \ref{fig:QO}.
	While the amplitude $|Q|$ is identical with the vacuum state, the quasi\-state characteristics are clearly visible through the nontrivial phases, $\arg Q\neq0$.

\subsubsection{Non-$s$-parametrized quasi\-states}

	Moreover, we can also consider the quasi\-states for cases with $p,q\neq0$.
	As we demand $|q|=|p|$, we have two extremal scenarios, $q=p^\ast$ and $q=-p^\ast$.

	In addition to the $s$-parametrized quasiprobabilities in Table \ref{tab:sParameters}, we also listed the Agarwal-Wolf distributions.
	They correspond to the case $q=-p^\ast=\mp 1$ (also, $s=0$).
	The complex-valued $Q$ function of the resulting quasi\-state $\hat\Delta_{F^{-1}\Omega F}$ is shown in the bottom-right plot of Fig. \ref{fig:QO}.
	The amplitude $|Q|$ is again compatible with the uncertainty principle bounded by the vacuum state, similarly to the previously discussed scenario of a complex $s$ parameter.
	Here, however, the functional behavior of the phase $\arg Q$ does not possess a radial symmetry.
	Rather, the quasi\-states to the Agarwal-Wolf distributions exhibit hyperbolic isophase contours.
	Note that this dismisses a sometimes held believe that Agarwal-Wolf distributions are an example of a $s$-parametrized quasiprobability.

	Finally, we can ask ourselves what happens in the case $q=p^\ast$.
	An example is shown in the top-right plot in Fig. \ref{fig:QO}.
	Interestingly, the resulting quasi\-states are squeezed.
	Specifically for a squeezing parameter $\zeta$, we can choose
	\begin{align}
		s=\frac{1+\tanh^2|\zeta|}{1-\tanh^2|\zeta|}
		\text{ and }
		q=\frac{-2e^{i\arg\zeta}\tanh|\zeta|}{1-\tanh^2|\zeta|}
		=p^\ast.
	\end{align}
	This choice implies $\bar n=0$ and means that the quasi\-states are pure squeezed states, cf. Eqs. \eqref{eq:ParameterRelations} and \eqref{eq:ParameterRelationsPlus}.
	Consequently, the quasiprobability density $P_{F^{-1}\Omega F}(\alpha)$ now describes the decomposition of the density operator in terms of displaced squeezed states with a coherent amplitude $\alpha$ and a squeezing defined by $\zeta$.
	This further implies that we have a generalized Glauber-Sudarshan quasiprobability distribution which is, however, a phase-space representation that expands the state in terms of squeezed states.
	Thus, the case $q=p^\ast$ significantly extends the notion of $s$-parametrized distributions to additionally include arbitrary squeezed states.

\section{Quantum correlations}\label{sec:Correlations}

	In a recent work \cite{SW18}, we generalized the concept of quasiprobability representations to more general notions of quantum coherence \cite{SAP17}, beyond the specific example of harmonic oscillators studied in the previous section.
	Among the various types of quantumness, quantum correlations between multiple degrees of freedom play an outstanding role for applications in quantum technologies \cite{NC00,HHHH09}.
	For this reason, let us study the entanglement of quantum systems within the framework of quasi\-states.
	A comprehensive introduction to the theory of entanglement (likewise, inseparability) can be found in Ref. \cite{HHHH09}.

\subsection{Entanglement quasiprobabilities}

	A pure separable state in a multipartite system is described by a tensor-product vector,
	\begin{align}
		|(a,b,\ldots)\rangle=|a\rangle\otimes|b\rangle\otimes\cdots,
	\end{align}
	where $a\in\mathcal S_A$, $b\in\mathcal S_B$, etc.
	In addition, the inclusion of classical correlations in terms of statistical mixtures, given by a classical probability density $P$, yields the definition of a mixed separable state \cite{W89},
	\begin{align}
		\label{eq:SepState}
		\hat\rho{=}\int_{S_A{\times} S_B{\times}\cdots} da\,db\cdots P(a,b,\ldots)|(a,b,\ldots)\rangle\langle (a,b,\ldots)|.
	\end{align}

	A state is defined to be inseparable (i.e., entangled) if it cannot be written in the form of Eq. \eqref{eq:SepState}.
	However, when we allow for $P$ to be a quasiprobability, even entangled states can be expanded in terms of separable states using Eq. \eqref{eq:SepState} \cite{STV98}.
	A construction approach for bipartite entanglement quasiprobabilities was introduced in Ref. \cite{SV09}.
	As this approach can be generalized from the bipartite scenario to the multipartite case \cite{SW18}, we restrict ourselves to bipartite systems in the following.

	The optimal decomposition of entangled states in terms of separable ones warrants that the entanglement is uniquely identified through negativities in $P(a,b)$ \cite{SV09,SW18}.
	The separable states $|(a,b)\rangle$ needed for the decomposition are obtained from the solution of the so-called separability eigenvalue equations (SEEs),
	\begin{align}
		\label{eq:SEEs}
		\hat\rho_a|b\rangle=g|b\rangle
		\text{ and }
		\hat\rho_b|a\rangle=g|a\rangle,
	\end{align}
	with $\hat\rho_a=\mathrm{tr}_A[\hat\rho(|a\rangle\langle a|\otimes \hat 1_B)]$ and $\hat\rho_b=\mathrm{tr}_B[\hat\rho(\hat 1_A\otimes|b\rangle\langle b|)]$.
	The SEEs approach generalizes the eigenvalue problem to composite systems while respecting the tensor-product structure of the corresponding eigenvectors and was initially introduced to construct entanglement witnesses \cite{SV09witn,SV13}.
	The solutions $(a,b)\in\mathcal S$ allow us to construct the distribution $P$ by solving the linear problem
	\begin{align}
		G\vec p=\vec g,
	\end{align}
	where $G=[|\langle (a,b)|(a',b')\rangle|^2]_{(a,b),(a',b')\in\mathcal S}$ is the Gram-Schmidt matrix of separability eigenvectors.
	Furthermore, the solution vector $\vec p=[P(a,b)]_{(a,b)\in\mathcal S}$ defines the quasiprobabilities, and $\vec g=[g]_{(a,b)\in\mathcal S}$ is the vector of separability eigenvalues, cf. Eq. \eqref{eq:SEEs}.
	Let us point out that the here-used Gram-Schmidt matrix also relates to the reconstruction approach in Sec. \ref{sec:StateRecon} [Eq. \eqref{eq:GramSchmidt}] and is a result of a underlying principle of convex decomposition, discussed in greater detail in Ref. \cite{SW18}.

	One can always find a set $\mathcal S_0$ such that $\mathcal S\subset\mathcal S_0\times \mathcal S_0$; we then set $P(a,b)=0$ for all $(a,b)\in \mathcal S_0\times \mathcal S_0\setminus\mathcal S$.
	Using the construction to find optimal entanglement quasiprobabilities via the SEEs, we can decompose any state, be it entangled or separable \cite{SV09,SW18}, as
	\begin{align}
		\label{eq:InitialBiState}
		\hat\rho=\sum_{(a,b)\in\mathcal S_0\times\mathcal S_0}P(a,b)|(a,b)\rangle\langle (a,b)|.
	\end{align}
	It is worth mentioning that the entanglement quasiprobability is always real valued and normalized.

\subsection{Quasi\-state representation}

	A universally applicable approach to diminish the negativities in quasiprobabilities for a single system is a convolution of the form
	\begin{align}
		P_K(\chi)=rP(\chi)+\frac{1-r}{|\mathcal S_0|}\sum_{\psi\in\mathcal S_0}P(\psi),
	\end{align}
	with $0< r\leq 1$ and $|\mathcal S_0|$ being the cardinality of the set; see Appendix \ref{app:UniformMix} for details.
	The kernel under study consequently leads to quasi\-states of the form
	\begin{align}
		\label{eq:UniformQuasistates}
		\hat\Delta_K(\chi)=\frac{1}{r}|\chi\rangle\langle\chi|-\frac{1-r}{r|\mathcal S_0|}\sum_{\psi\in\mathcal S_0}|\psi\rangle\langle\psi|.
	\end{align}
	We emphasize that quasi\-states defined in this manner are both Hermitian, $\hat\Delta_K(\chi)=\hat\Delta_K(\chi)^\dag$, and normalized, $\mathrm{tr}[\hat\Delta_K(\chi)]=1$.
	Thus, the only distinction to true states is the fact that $\hat\Delta_K(\chi)$ is, in general, not a positive semidefinite operator.

	The convolution can be generalized to composite systems via product kernels.
	Choosing, for example, the same $r$ for both systems, we can reformulate Eq. \eqref{eq:InitialBiState} as
	\begin{align}
		\label{eq:ModifiedBiState}
		\hat\rho=\sum_{(\tilde a,\tilde b)\in\mathcal S_0\times\mathcal S_0}P_{K\otimes K}(\tilde a,\tilde b)\,\hat\Delta_K(\tilde a)\otimes \hat\Delta_K(\tilde b).
	\end{align}
	In particular, we have
	\begin{align}
		\label{eq:BipartConv}
	\begin{aligned}
		P_{K\otimes K}(\tilde a,\tilde b)
		=&r^2P(\tilde a,\tilde b)+\frac{(1-r)^2}{|\mathcal S_0|^2}
		\\
		&+\frac{r(1-r)}{|\mathcal S_0|}\left[P(\tilde a)+P(\tilde b)\right],
	\end{aligned}
	\end{align}
	using the marginal distributions $P(\tilde a)=\sum_{b\in\mathcal S_0}P(\tilde a,b)$ and $P(\tilde b)=\sum_{a\in\mathcal S_0}P(a,\tilde b)$.

	Eventually, we get a nonnegative distribution $P_{K\otimes K}$ for a sufficiently large $r$, which is also demonstrated later.
	This means that an entangled state can be decomposed according to Eq. \eqref{eq:ModifiedBiState} in terms of a classical joint probability distribution $P_{K\otimes K}(a,b)\geq0$ and tensor-product quasi\-states $\hat\Delta_K(a)\otimes \hat\Delta_K(b)\ngeq0$ [cf. Eq. \eqref{eq:UniformQuasistates}].
	Since we can chose $r=0$ for separable states, we find that quasi\-states are not required in this case.
	Therefore, we are able to conclude that entanglement is unambiguously identified in terms of classical joint probability distributions if and only if the tensor-product operators of the decomposition have to be unphysical quasi\-states.

\subsection{Example}

	As a proof of concept, let us consider an entangled two-qubit quantum state parametrized as
	\begin{align}
		\label{eq:TwoQubit}
		\hat\rho=\frac{1}{4}\left[
			\hat 1\otimes\hat 1
			+\sum_{j\in\{x,y,z\}}\rho(j)\,\hat \sigma_j\otimes\hat \sigma_j
		\right].
	\end{align}
	This corresponds to a physical density operator if and only if the real coefficients satisfy $[\rho(x),\rho(y),\rho(z)]\in\mathrm{conv}\{[-1,1,1],[1,-1,1],[1,1,-1],[-1,-1,-1]\}$, where the extremal points represent Bell states.
	In Ref. \cite{SW18}, we solved the SEEs [Eq. \eqref{eq:SEEs}] for this family of states.
	In particular, the separability eigenvectors are tensor products of eigenstates of Pauli operators ($\hat\sigma_x$, $\hat\sigma_y$, and $\hat\sigma_z$).
	This implies that we have
	\begin{align}
		\label{eq:TwoQubitSet}
		\mathcal S_0=\{
			x_{+},x_{-},y_{+},y_{-},z_{+},z_{-}
		\},
	\end{align}
	where $\hat\sigma_w|w_s\rangle=s|w_{s}\rangle$ for $w_s\in\mathcal S_0$.

	The exact entanglement quasiprobability reads \cite{SW18}
	\begin{align}
		\label{eq:QuasiProbTwoQubit}
		P(a_s,b_t)=\delta(a,b)\left[
			\frac{q}{12}
			+\frac{|\rho(a)|+st\rho(a)}{4}
		\right],
	\end{align}
	with $a,b\in\{x,y,z\}$ and $s,t\in\{+1,-1\}$, identifying the separability eigenvectors, $(a_s, b_t)\in\mathcal S_0\times\mathcal S_0$.
	Of particular importance is the parameter
	\begin{align}
		q=1-|\rho(x)|-|\rho(y)|-|\rho(z)|
	\end{align}
	as it determines if the state is separable; namely, $\hat\rho$ in Eq. \eqref{eq:TwoQubit} is separable if and only if $q\geq 0$ \cite{SW18}.
	Conversely, the negativity of the quasiprobability in the case of entanglement is given by this quantity as well; it reads $\min_{(u, v)\in\mathcal S_0\times\mathcal S_0}P(u,v)=q/12\geq-1/6$, and the lower bound is attained for Bell states.

\begin{figure}[b]
	\includegraphics[width=.8\columnwidth]{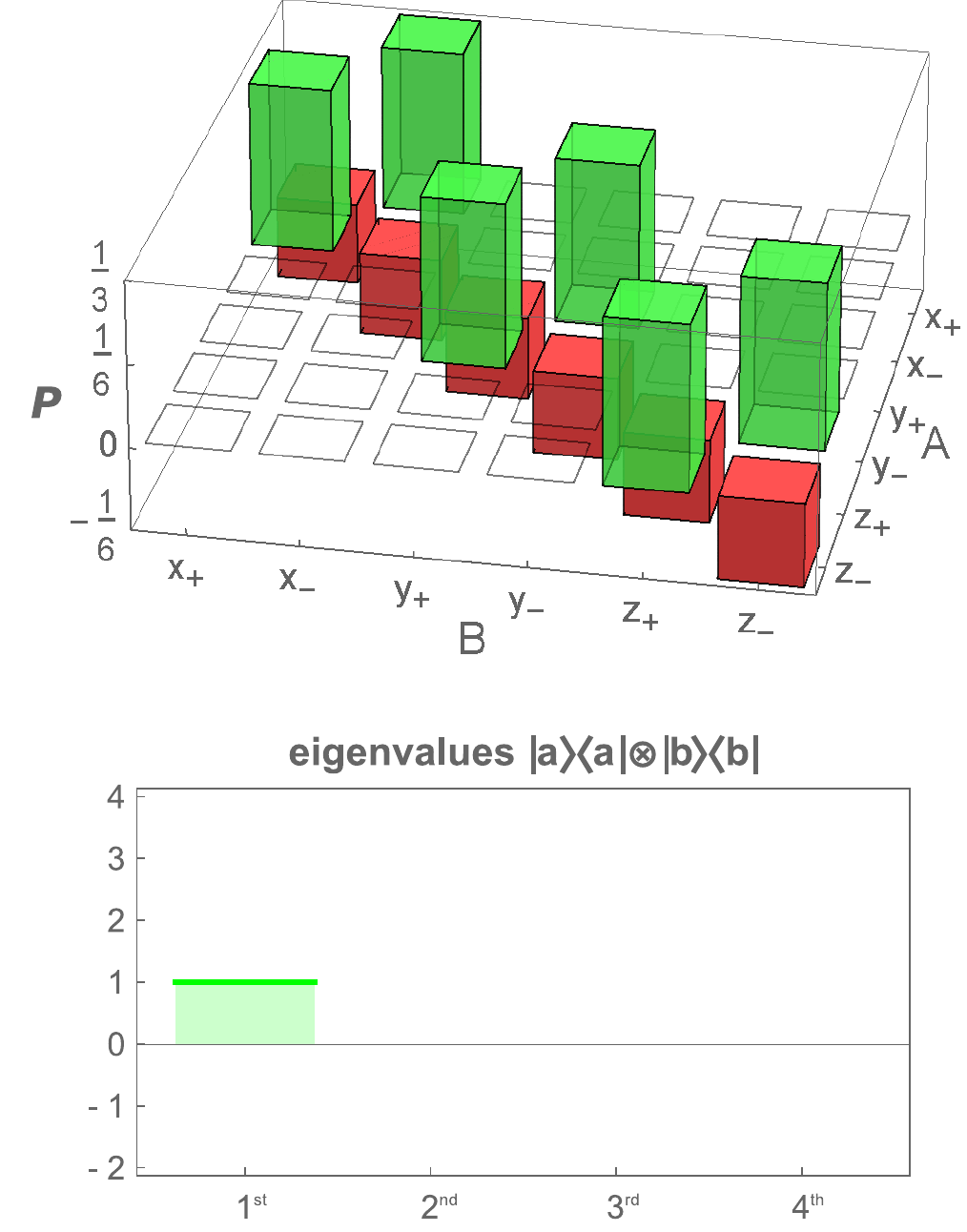}
	\caption{(Color online)
		The top panel depicts the entanglement quasiprobabilities [Eq. \eqref{eq:QuasiProbTwoQubit}] for a Bell state with $\rho(x)=\rho(y)=\rho(z)=-1$.
		Each patch in the $A$-$B$ plane corresponds to a product state $|a\rangle\langle a|\otimes|b\rangle\langle b|$, with $a,b\in\mathcal S_0$ according to Eq. \eqref{eq:TwoQubitSet}.
		The bottom panel shows the eigenvalues of those separable-state projectors.
		The negativities in the quasiprobabilities $P$ identify the entanglement.
	}\label{fig:entanglement1}
\end{figure}

	Now we can apply the kernel in Eq. \eqref{eq:BipartConv}.
	Inserting the distribution in Eq. \eqref{eq:QuasiProbTwoQubit} yields
	\begin{align}
		\label{eq:FilteredProbTwoQubit}
	\begin{aligned}
		&P_{K\otimes K}(a_s,b_t)
		\\
		= & r^2\delta(a,b)\left[
			\frac{q}{12}
			+\frac{|\rho(a)|+st\rho(a)}{4}
		\right]
		+\frac{(1-r)^2}{36}
		\\
		&+\frac{r(1-r)}{6}\left[
			\frac{q}{3}+\frac{|\rho(a)|}{2}+\frac{|\rho(b)|}{2}
		\right].
	\end{aligned}
	\end{align}
	After some straightforward algebra, we find that $P_{K\otimes K}$ is always nonnegative for $0<r\leq 1/\sqrt{7}$.
	In addition, we get tensor-product quasi\-states which are formed by the subsystem components
	\begin{align}
		\label{eq:2QubitQuasistates}
		\hat\Delta_{K}(j)=\frac{1}{r}|j\rangle\langle j|-\frac{1-r}{2r}\hat 1
	\end{align}
	for $j\in\mathcal S_0$ (see Appendix \ref{app:UniformMix}).
	Consequently, the eigenvalues of $\hat\Delta_{K}(j)$ are $(1+r)/(2r)$ and $-(1-r)/(2r)$.

\begin{figure}[tb]
	\includegraphics[width=.8\columnwidth]{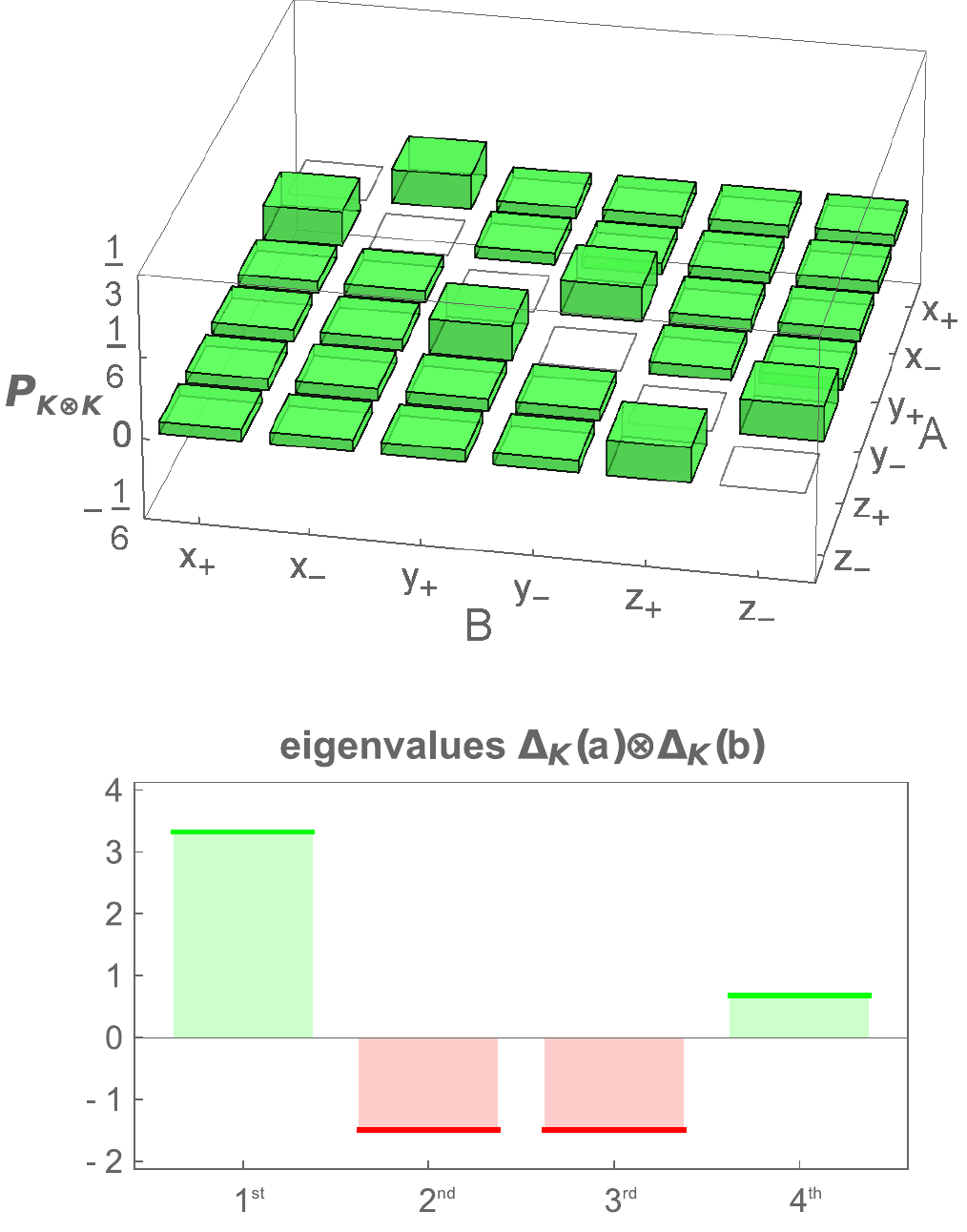}
	\caption{(Color online)
		The top panel depicts the classical joint probabilities, obtained using the kernel $K$ [see Eq. \eqref{eq:FilteredProbTwoQubit} for $r=1/\sqrt7$], for the same Bell state as shown in Fig. \ref{fig:entanglement1}.
		The bottom panel shows the positive and negative eigenvalues of the quasi\-states in Eq. \eqref{eq:2QubitQuasistates} which are then required for the decomposition of the entangled state.
		Now the entanglement is captured by the unphysical nature of the quasi\-states.
	}\label{fig:entanglement2}
\end{figure}

	Let us consider a Bell state $\hat\rho=|\psi\rangle\langle\psi|$ as a specific example, where
	\begin{align}
		|\psi\rangle=\frac{|0\rangle\otimes|1\rangle-|1\rangle\otimes|0\rangle}{\sqrt 2},
	\end{align}
	i.e., $\rho(x)=\rho(y)=\rho(z)=-1$.
	The quasiprobability decomposition of this state is shown in Fig. \ref{fig:entanglement1}.
	The negativities in the joint quasiprobability, $P\ngeq0$, certify the entanglement and enable us to decompose the entangled state $\hat \rho$ using nonnegative tensor-product projectors, $|a\rangle\langle a|\otimes|b\rangle\langle b|\geq0$.
	In contrast, in Fig. \ref{fig:entanglement2}, we show the result of applying the uncorrelated and nonnegative kernel $K\otimes K$ for $r=1/\sqrt7$; see Eq. \eqref{eq:FilteredProbTwoQubit}.
	This results in a joint classical probability distribution.
	However, the consequence is that the tensor-product operators for the decomposition have to be unphysical quasi\-states (because of the negative eigenvalues) in order to expand the entangled Bell state under study.

\section{Conclusions and discussion}\label{sec:SumConcl}

	Since the early theoretical descriptions of quantum states in phase space, quasiprobabilities have become one of the most important tools for studying quantum phenomena.
	Here, we complemented this treatment by introducing the concept of quasi\-states.
	While nonclassical states can be expanded in terms of a quasiprobability density and physical states, we demonstrated that one can equivalently use a classical, i.e., nonnegative, distribution and quasi\-states to perform such an expansion.
	Then the quantum features of the state are carried over from the necessity of negativities in the distribution to the requirement that unphysical quasi\-states are needed to describe the state of a system.
	Furthermore, we elaborated a number of different aspects and applications of quasi\-state, rendering them a useful tool to characterize quantum systems.

	In simple terms, quasi\-states almost describe a density operator---only certain properties of a physical state do not apply.
	In particular, a physical density operator is a Hermitian and positive semidefinite operator with a unit trace.
	Throughout this work, we discussed several examples for which one (or multiple) of the defining properties of a density operator are violated by a quasi\-state.
	For instance, eigenvalues can be negative, violating the positive semidefiniteness, or a Fourier representation was shown to lead to non-Hermitian quasi\-states.

	As one practical implementation, we studied the reconstruction of the density operator.
	This was based on the findings of a recent work \cite{KSVS18}, where Born's rule was reinterpreted in terms of contravariant operator-valued measures.
	For example, this dual concept is useful when the positive operator-valued measure describes imperfect measurement outcomes.
	Here, we proved that the contravariant operator-valued measure directly relates to the concept of quasi\-states and, therefore, can be considered as a special case of our general notion.
	Consequently, we were able to apply quasi\-states for a direct quantum state reconstruction in terms of measured probabilities, such as exemplified for a two-level qubit state.

	The duality between quasiprobability densities and their corresponding quasi\-states was studied for prominent quantum-optical phase-space distributions.
	For instance, we found that the quasi\-states of the Wigner-Weyl distribution coincide with an operator that describes the maximal singularities the Glauber-Sudarshan distribution can have \cite{S16}.
	Beyond known quasiprobabilities in quantum optics, we further showed that our approach also leads to generalized phase-space distributions which expand the state in terms of squeezed states rather than coherent ones.
	This, for example, can be useful to characterize non-Gaussianity, which is required for universal quantum computation and, in contrast to nonclassicality in terms of the Glauber-Sudarshan distribution, not only relies on coherent states but general squeezed states \cite{KV18}.

	Furthermore, we characterized quantum correlations using quasi\-states.
	In particular, we showed that entanglement can be identified with a classical joint probability distribution and tensor-product quasi\-states.
	This result is surprising when considering that a violation of local-hidden-variable models typically excludes such classical distributions because states are (obviously) implicitly assumed to be physical in such models.
	Here, we proved the necessary and sufficient condition that a density operator is separable, i.e., not entangled, if and only if both the distribution and the used states are classical; conversely, entanglement requires that at least one of them is nonclassical.
	As an example, we identified the entanglement of a Bell state using either optimal quasiprobabilities \cite{SV09} and tensor-product states or classical probabilities and unphysical quasi\-states.
	Based on the general construction of quasiprobabilities for other notions of quantum coherences \cite{SW18}, the found results can be straightforwardly generalized to other forms of quantumness.

	In conclusion, we developed the versatile and useful framework of quasi\-states for decomposing density operators.
	Previously established concepts and methods have been demonstrated to be equivalently accessible with our technique which additionally allowed us to go beyond this state of the art.
	Thus, we believe that, analogously to quasiprobabilities, the notion of quasi\-states has the potential to significantly contribute to the description and reconstruction of nonclassical quantum states in theory and experiment.

\begin{acknowledgments}
	This work has received funding from the European Union's Horizon 2020 Research and Innovation Program under Grant Agreement No. 665148 (QCUMbER).
\end{acknowledgments}

\appendix

\section{Exponential functions of bosonic operators}\label{app:ExpOp}

	For our treatment of quantum light in Sec. \ref{sec:NclLight}, some additional algebra is required.
	In this appendix, we formulate the needed relations.

\subsection{General relations}

	We frequently apply the Gaussian integral formula
	\begin{align}
		\label{eq:IntegralFormula}
	\begin{aligned}
		&\int_{\mathbb C}d\xi\,\exp\left(
			-z|\xi|^2-x\xi^2-y\xi^{\ast2}
			+u\xi+v\xi^\ast
		\right)
		\\
		=&\frac{\pi}{\sqrt{z^2-4xy}}
		\exp\left(\frac{
			zuv-yu^2-xv^2
		}{z^2-4xy}\right),
	\end{aligned}
	\end{align}
	where $\mathrm{Re}[z\pm(x+y)]>0$ and $\mathrm{Re}[z^2-4xy]>0$.
	We use the branch with a nonnegative real part for the square root of a complex number.
	The above relation can be extended to integrals of normally ordered expressions because under this prescription, we have an algebra of commuting operators.

	In the case that operator ordering becomes relevant, let us formulate additional relations.
	From the observations that $\hat a^mf(\hat n)|n\rangle=f(\hat n+m)\hat a^m|n\rangle$ and $f(\hat n)\hat a^{\dag m}|n\rangle=\hat a^{\dag m}f(\hat n+m)|n\rangle$ hold true, we conclude that
	\begin{align}
		\label{eq:CommuteExp1}
		e^{x\hat a^2}\omega^{\hat n}=\omega^{\hat n}e^{\omega^2x\hat a^2}
		\text{ and }
		\omega^{\hat n}e^{y\hat a^{\dag 2}}=e^{\omega^2y\hat a^{\dag 2}}\omega^{\hat n}.
	\end{align}
	In the same manner, we get $\exp(u\hat a)\omega^{\hat n}=\omega^{\hat n}\exp(\omega u\hat a)$ and $\omega^{\hat n}\exp(v\hat a^\dag)=\exp(\omega v\hat a^\dag)\omega^{\hat n}$.
	Furthermore, using the above integral, $|\alpha\rangle\langle\alpha|={:}\exp(-[\hat a-\alpha]^\dag[\hat a-\alpha]){:}$, and $\pi\hat 1=\int_{\mathbb C}d\alpha\,|\alpha\rangle\langle\alpha|$, we obtain
	\begin{align}
		\label{eq:CommuteExp2}
	\begin{aligned}
		&e^{x\hat a^2}e^{y\hat a^{\dag2}}
		=\exp[x\hat a^2]\hat 1\exp[y\hat a^{\dag2}]
		\\
		=&\frac{1}{\sqrt{1-4xy}}
		{:}\exp\left(\frac{4xy\hat a^\dag\hat a+x\hat a^2+y\hat a^{\dag2}}{1-4uv}\right){:}
		\\
		=&
		e^{v\hat a^{\dag 2}/(1-4uv)}
		\left(\frac{1}{1-4uv}\right)^{\hat n+1/2}
		e^{u\hat a^2/(1-4uv)}.
	\end{aligned}
	\end{align}
	The same approach yields $\exp[u\hat a]\exp[y\hat a^{\dag2}]=\exp[y(\hat a^\dag+u)^2]\exp[u\hat a]$ and $\exp[x\hat a^2]\exp[v\hat a^\dag]=\exp[v\hat a^\dag]\exp[x(\hat a+v)^2]$.
	Finally, let us also mention the well-known formula $\exp(u\hat a)\exp(v\hat a^\dag)=\exp(uv)\exp(v\hat a^\dag)\exp(u\hat a)$.

\subsection{Spectral decomposition of Gaussian quasi\-states}

	In the main part of this contribution, we consider filter functions of a Gaussian form [cf. Eq. \eqref{eq:GaussFilter}].
	The exact analysis for the resulting quasi\-states is performed here.
	For zero displacement, the integral in Eq. \eqref{eq:IntegralFormula} yields
	\begin{align}
		\label{eq:AppQuasiState}
	\begin{aligned}
		\hat\Delta
		=&\int_{\mathbb C}d\beta\,\frac{e^{-|\beta|^2}}{\pi\Omega(\beta)}e^{-\beta\hat a^\dag}e^{\beta^\ast\hat a}
		\\
		=&\frac{2\,{:}\exp\left(
			\frac{-2[1+s]\hat a^\dag\hat a +q\hat a^{\dag 2}+p\hat a^2}{(1+s)^2-pq}
		\right){:}}{\sqrt{(1+s)^2-pq}}.
	\end{aligned}
	\end{align}

	To characterize this quasi\-state, let us consider its decomposition in terms of exponential operators.
	First, the squeezing operator $\hat S=\exp([\zeta^\ast\hat a^2-\zeta\hat a^{\dag2}]/2)$ can be equivalently given as \cite{VW06}
	\begin{align}
		\hat S=
		\exp\left(
			-\frac{\tau}{2}\hat a^{\dag2}
		\right)
		\sqrt{1-|\tau|^2}^{\hat n+1/2}
		\exp\left(
			\frac{\tau^\ast}{2}\hat a^2
		\right),
	\end{align}
	where $\tau=e^{i\arg\zeta}\tanh|\zeta|$.
	Note that $|\tau|<1$, and we have infinite squeezing for $|\tau|\to1$.
	In addition, we consider thermal-state-like operator
	\begin{align}
	\begin{aligned}
		\hat W=&(1-\omega)\omega^{\hat n}
		=\sum_{n\in\mathbb N}(1-\omega)\omega^{n}|n\rangle\langle n|
		\\
		=&(1-\omega){:}e^{-(1-\omega)\hat n}{:}
		\\
		=&\int_{\mathbb C}d\alpha\,\frac{1-\omega}{\pi\omega}
		\exp\left(
			-\frac{(1-\omega)|\alpha|^2}{\omega}
		\right)
		|\alpha\rangle\langle\alpha|.
	\end{aligned}
	\end{align}
	See Ref. \cite{SVA14} for a derivation and additional considerations.
	Also note that the vacuum state $\hat W=|0\rangle\langle 0|$ is obtained in the limit $\omega\to0$.

	Now, we defined an operator $\hat T$ to be compared with the quasi\-state $\hat \Delta$,
	\begin{align}
		\hat T=\hat S\hat W\hat S^\dag.
	\end{align}
	Using the exchange relations in Eqs. \eqref{eq:CommuteExp1} and \eqref{eq:CommuteExp2}, we find after some algebra a normally ordered expression,
	\begin{align}
	\begin{aligned}
		\hat T
		=&(1-\omega)\sqrt{\frac{
			1-|\tau|^2
		}{
			1-|\tau|^2\omega^2
		}}
		\\
		&\times
		\exp\left(
			-\frac{\tau}{2}\frac{1-\omega^2}{1-|\tau|^2\omega^2}\hat a^{\dag2}
		\right)
		\\
		&\times
		{:}\exp\left(
			-\frac{[1-\omega][1+\omega|\tau|^2]}{1-|\tau|^2\omega^2}\hat n
		\right){:}
		\\
		&\times
		\exp\left(
			-\frac{\tau^\ast}{2}\frac{1-\omega^2}{1-|\tau|^2\omega^2}\hat a^2
		\right).
	\end{aligned}
	\end{align}
	Finally, we can compare this expression with Eq. \eqref{eq:AppQuasiState}, $\hat\Delta=\hat T$.
	Namely, equating coefficient yields
	\begin{align}
		\omega=\frac{r-1}{r+1},\,
		|\tau|^2=\frac{s-r}{s+r},\,
		\text{and}\,
		\frac{\tau}{\tau^\ast}=\frac{q}{p},
	\end{align}
	where $r=\sqrt{s^2-pq}$.
	This also implies $\tau=-q/(s+r)$ and $\tau^\ast=-p/(s+r)$.
	Conversely, we can deduce parameters $s$, $p$, and $q$ from given values of $\omega$ and $\tau$,
	\begin{align}
	\begin{aligned}
		p=\frac{-2(1+\omega)\tau^\ast}{(1-\omega)(1-|\tau|^2)},
		\quad
		q=\frac{-2(1+\omega)\tau}{(1-\omega)(1-|\tau|^2)},
		\\
		\text{and}\quad
		s+1=\frac{2(1+\omega|\tau|^2)}{(1-\omega)(1-|\tau|^2)}.
	\end{aligned}
	\end{align}

\section{Uniform attenuation}\label{app:UniformMix}

	In the main text, we consider a specific example of a kernel.
	For studying its properties, let us consider the corresponding convolution
	\begin{align}
		P_K(\chi)=\sum_{\psi\in\mathcal S}P(\psi)K(\psi,\chi)=a\,P(\chi)+b\sum_{\psi\in\mathcal S}P(\psi)
	\end{align}
	for a set $\mathcal S$ of states.
	Note that $\sum_{\psi\in\mathcal S}P(\psi)=1$ is the normalization condition.
	To ensure that the result is normalized as well, $1=\sum_{\chi\in\mathcal S}P_K(\chi)=a+b|\mathcal S|$, we find
	\begin{align}
		b=\frac{1-a}{|\mathcal S|},
	\end{align}
	using the cardinality $|\mathcal S|$ of the set of states.
	Further, when $0\leq a\leq 1$ is satisfied, the non-negativity of $P_K$ is guaranteed for any nonnegative $P$.
	The considered convolution can be formulated via the $|\mathcal S|\times|\mathcal S|$ map
	\begin{align}
		K=a\,\mathrm{id}+b\,nn^\mathrm{T},
	\end{align}
	using the identity $\mathrm{id}$ and the $|\mathcal S|$-dimensional vector $n=[1,\ldots,1]^\mathrm{T}$.
	Then the inverse takes a similar form,
	\begin{align}
		K^{-1}=\frac{1}{a}\mathrm{id}-\frac{b/a}{a+b|\mathcal S|}nn^\mathrm{T},
	\end{align}
	where $a+b|\mathcal S|=1$ for the given $b$.
	Using the normalization, the convolution kernel $K$ corresponds to uniform addition of noise to the probabilities.
	On the level of the operators, we get
	\begin{align}
		\hat\Delta_K(\chi)=\frac{1}{a}|\chi\rangle\langle\chi|-\frac{1-a}{a|\mathcal S|}\sum_{\psi\in\mathcal S}|\psi\rangle\langle\psi|.
	\end{align}

	In the continuous case, we can generalize the above relations as
	\begin{align}
		P_K(\chi)=\int_{\mathcal S} d\psi\,P(\psi)\left[a\delta(\psi,\chi)+\frac{1-a}{|\mathcal S|}\right],
	\end{align}
	where the volume $|\mathcal S|=\int_{\mathcal S}d\psi$ is used instead.
	Furthermore, the projection operators transform as
	\begin{align}
		\hat\Delta_{K}(\chi)=\int_{\mathcal S} d\psi\left[\frac{1}{a}\delta(\psi,\chi)-\frac{1-a}{a|\mathcal S|}\right]|\psi\rangle\langle\psi|.
	\end{align}

	As a discrete-variable example, we study a qubit system with orthogonal basis $\{|0\rangle,|1\rangle\}$.
	We can identify the Pauli matrices as $\hat\sigma_z=|1\rangle\langle1|-|0\rangle\langle0|$, $\hat\sigma_x=|0\rangle\langle1|+|1\rangle\langle0|$, and $\hat\sigma_y=i|0\rangle\langle1|-i|1\rangle\langle0|$.
	The Hermitian operator basis can be completed with the operator $\hat 1=|0\rangle\langle 0|+|1\rangle\langle 1|$.
	The eigenvectors $|j_{\pm1}\rangle$ of the Pauli matrices $\hat\sigma_j$ to the eigenvalues $\pm1$ for $j\in\{z,x,y\}$ form the set $\mathcal S$.
	Then we can decompose any density operator as
	\begin{align}
		\hat\rho=\sum_{j\in\{x,y,z\},s\in\{+1,-1\}}P(j_s)|j_{s}\rangle\langle j_{s}|.
	\end{align}
	Conversely, the attenuated distribution reads $P_{K}(j_s)=aP(j_s)+(1-a)/6$, and the transformed operators take the form $\hat\Delta_K(j_s)=a^{-1}|j_s\rangle\langle j_s|-(1-a)(2a)^{-1}\hat 1$.


\end{document}